\documentclass[
 amsmath,amssymb,
 aps,physrev,
 prevappl, 
twocolumn
]{revtex4-2}

\usepackage{graphicx}
\usepackage{dcolumn}
\usepackage{bm}
\usepackage{hyperref}
\usepackage{soul} 
\usepackage{xcolor}

\begin{document}


\title{\textbf{ Magnetic Particle Magnetometer Based on Langevin Response of Superparamagnetic Nanoparticles for Remote Magnetic Field Sensing} 
}%
\author{M. Jurj}
\affiliation{Physics Department, Oakland University, Rochester, MI 48309 USA}
\author{C. McDonough}
\affiliation{Physics Department, Oakland University, Rochester, MI 48309 USA}
\author{C. Bastajian}
\altaffiliation[Also at ]{Department of Electrical and Computer Engineering}
\author{A. Tonyushkin}
\email{tonyushkin@oakland.edu}%
\affiliation{Physics Department, Oakland University, Rochester, MI 48309 USA}

\date{\today}%

\begin{abstract}
We demonstrate a novel magnetometer that utilizes the Langevin nonlinear magnetization of superparamagnetic nanoparticles as the sensing medium. The magnetic sensor exploits the even-order harmonics of the magnetization response, particularly the second harmonic, which is proportional to the local magnetic field component. By remotely detecting the magnetization response, the magnetometer enables non-invasive, localized, and distributed magnetic field sensing without requiring wired connections, with a small sensing volume of $\sim \mu$l, which can be used for {\em in situ} biomedical sensing, geomagnetic monitoring, smart diagnostics, and distributed industrial sensing. In the linear operating regime, the magnetometer provides a measurement range of $\pm$ 200 $\mu$T and a high sensitivity of 10~pT/$\sqrt{\mathrm{Hz}}$, while an extended operating mode allows measurements over a range exceeding $\pm$ 0.5 mT. The sensor offers a dynamic bandwidth from dc up to a few kilohertz, determined by the excitation frequency. To validate performance, we accurately measured the vertical component of the local Earth’s magnetic field and found good agreement with measurements obtained using an atomic magnetometer.
\end{abstract}


\maketitle

\section{\label{sec:intro}Introduction} 
\begin{figure}[htb!]
    \includegraphics[width=\linewidth]{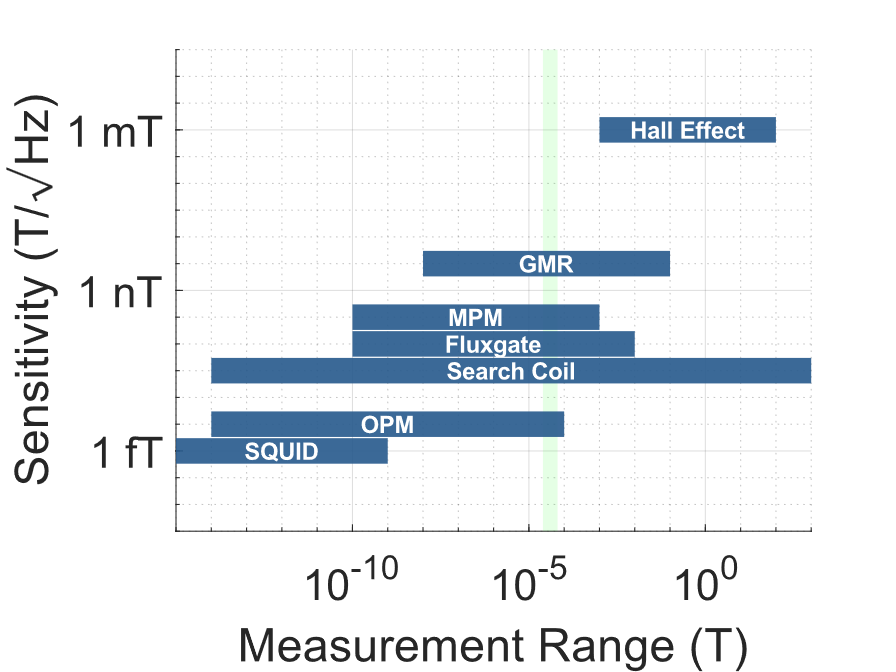}
    \caption{ Comparison of typical magnetometers' performance: the sensitivity vs. maximum range of the most common magnetometers. MPM measurement range is limited to the linear regime.}
    \label{compare}
\end{figure}

Magnetic field sensing remains fundamentally application-dependent, as no 
universal magnetometer simultaneously optimizes dynamic range, sensitivity, 
accuracy, stability, size, power consumption, and cost \cite{Lenz2006}. 
Existing magnetometer technologies span from compact solid-state devices to sophisticated quantum sensing platforms, each offering distinct advantages and limitations in sensitivity, dynamic range, bandwidth, and accuracy (see Fig.~\ref{compare}).
%
One of the most common highly sensitive magnetometers is based on magnetoresistance
(AMR, GMR, TMR) providing both sensitivity and miniaturization, but 
may be affected by drift and limited precision \cite{MAPPS1997, Zheng2019}. 
Fluxgate magnetometers offer excellent sensitivity, stability, and specificity, yet they have limited localization of the sensing region \cite{RIPKA1992, Janosek2017}. 
At the high-performance end, optically pumped magnetometers (OPMs) \cite{Budker2007} and superconducting quantum interference devices (SQUIDs) \cite{clarke2004} achieve exceptional sensitivity for biomagnetic and fundamental physics applications \cite{Jensen2016, Sternickel2006}, though both typically require highly controlled operating environments, with SQUIDs additionally depending on cryogenic cooling. Recently, a sensitive magnetometry method for biomedical applications was demonstrated using an OPM under ambient conditions \cite{Limes2020}.
The latest magnetic sensing technology, NV-diamond magnetometers, has emerged as a promising class of nanoscale quantum sensors capable of {\em in situ} magnetic measurements \cite{Taylor2008, Barry2016}, albeit at the expense of substantial complexity in microfabrication and in the optical and microwave systems. 

Despite significant advances in the above magnetic sensing technologies, existing 
magnetometers remain fundamentally constrained by the need for direct electrical 
connections, fixed sensor placement, or volume-averaged field measurements. As a 
result, truly non-invasive and distributed magnetic-field sensing in biological 
tissue or inaccessible environments remains largely unattainable. In particular, 
conventional sensors cannot be readily embedded into macroscopic living systems due to limitations in biocompatibility, size, and wiring requirements.

In this letter, we introduce a new type of harmonic magnetometer (MPM) in which superparamagnetic iron oxide nanoparticles (SPIONs) serve as a remotely interrogated sensing medium. Similar to a fluxgate magnetometer \cite{RIPKA1992}, the proposed approach exploits the generation of even-order harmonic responses, particularly the second harmonic, in a nonlinear medium, which serves as a specific spectral signature directly correlated with the local magnetic field. However, unlike conventional fluxgates that rely on saturation of a ferromagnetic core, the sensing element consists of remotely interrogated, hysteresis-free SPIONs. By remotely detecting these harmonics, the sensor can measure weak magnetic fields with high sensitivity and spatial specificity without physical contact or wired connections.  
Thus, the proposed sensor shares the robustness and specificity of fluxgates while incorporating advantages commonly associated with quantum sensors, including remote operation and embedded sensing capability, without the nanoscale fabrication complexity and cryogenic cooling. Since SPIONs are inherently biocompatible and naturally processed within the body \cite{Kolosnjaj-Tabi2016a}, including those used in existing FDA-approved tracers \cite{Stabi2011}, they provide a basis for embedded biomagnetic sensing. At excitations of $\sim$ 10-- 40 kHz, the odd harmonics of the magnetization response of functionalized SPIONs are widely used to detect and map concentration distributions {\em in vivo} using Magnetic Particle Imaging (MPI) \cite{Gleich2005, Panagiotopoulos2015} and Magnetic Particle Spectroscopy (MPS) \cite{Dhavalikar2015, Lowa2017} techniques. In the dc and low-frequency range, OPMs are frequently used to detect, image, and study the properties of SPIONs \cite{Xu2006,Colombo2016a,Colombo2020}. However, unlike those techniques, here we utilize SPIONs as a sensor of the external magnetic field {\em in situ}. 
Here, we provide a basic theory of MPM, demonstrate a proof-of-principle instrumentation, characterize its performance, and validate it with single-axis measurements of the Earth's magnetic field. 
\section{\label{sec:theory}Theory and Simulations}
\begin{figure*}[htb!]
    \includegraphics[width=\textwidth]{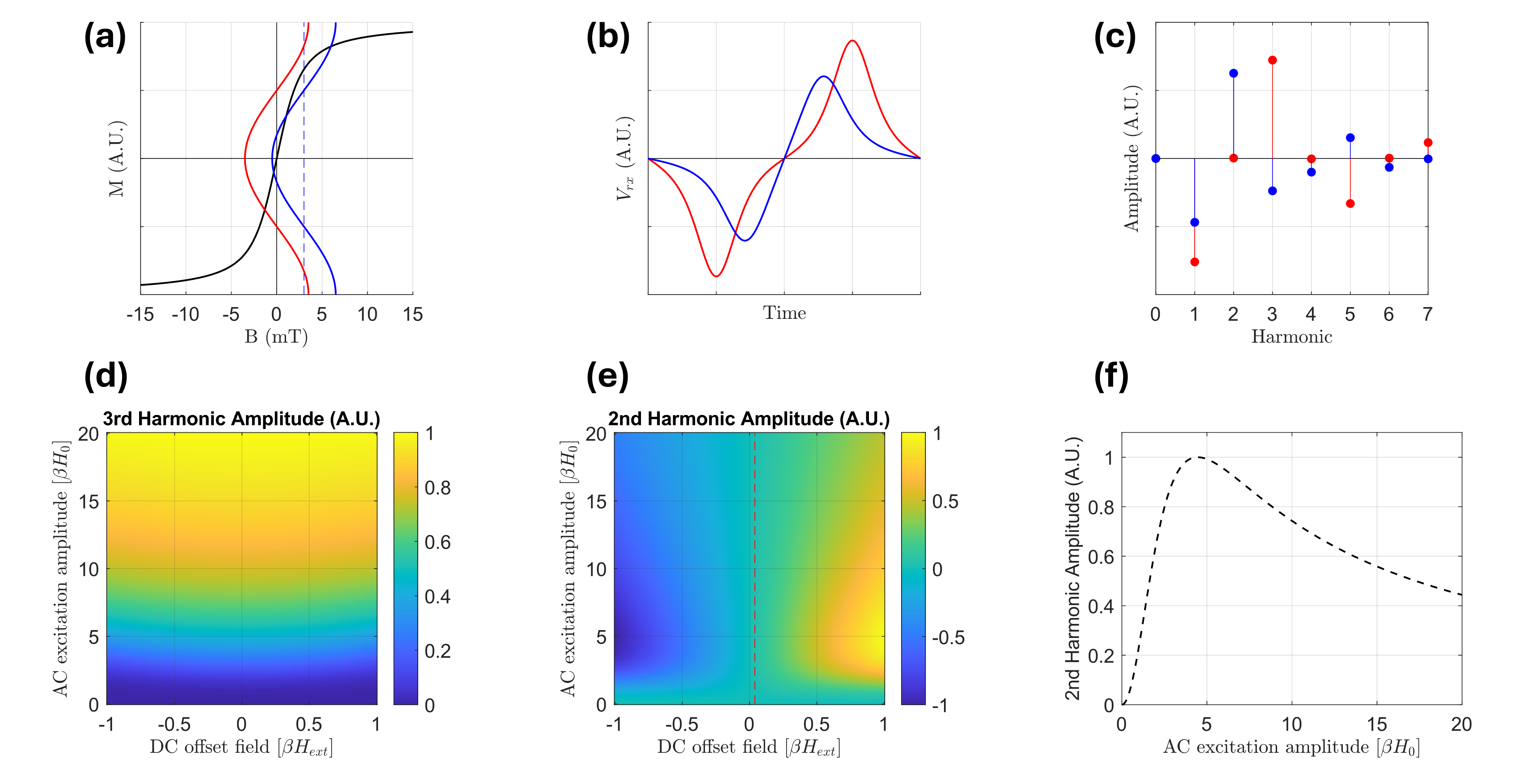}
    \caption{Simulated response of Langevin medium to an oscillating magnetic field. (a) Langevin magnetization curve (black) and total applied magnetic field with no dc offset (red) and with a dc offset field (blue). (b) Corresponding receive coil voltage signals due to Faraday induction. (c) Harmonic spectra of the receive coil voltage signals. (d) Third harmonic amplitude as a function of ac amplitude and dc offset. (e) Second harmonic amplitude as a function of ac amplitude and dc offset. (f) Second harmonic amplitude vs ac amplitude at set dc offset field $\beta H_{\rm ext}$.}
    \label{fig:theory}
\end{figure*}

The magnetometer measures the magnetic field based on the nonlinear magnetization 
curve in the superparamagnetic material. Such material as iron oxide in the form of Hematite or Magnetite, below the Curie temperature in a microscopic state with a size $\sim$10 nm, has a single-domain structure \cite{Elmore1938, Garcia1998}. 

When an SPION is placed in the external ac magnetic field $H(t)=H_0 sin(2\pi ft)$, the nonlinear magnetization response can be detected by a receive pickup (Rx) coil, which generates an {\it emf} via electromagnetic induction.

If the receive sensitivity is $\textbf{P}_{rx}$ then the {\it emf} due to field 
$\textbf{H}(\textbf{r},t)$ and $\textbf{M}(\textbf{r},t) $ is
\begin{equation}\label{rx_sig_eqn}
\begin{aligned}
  u(t) =&\, -\mu_0\frac{d}{d t}\int_{\Omega} \textbf{P}_{rx}(\textbf{r})\cdot \left\{ \textbf{H}(\textbf{r},t)+\textbf{M}(\textbf{r},t) \right\} dV \\
 & = -\mu_0 
 \int_{\Omega} \mathbf{P}_{rx}(\mathbf{r})\cdot \frac{\partial \mathbf{H}(\mathbf{r},t)}{\partial t} dV - \\
 & -\mu_0 \int_{\rm SPIO} c(\mathbf{r})\mathbf{P}_{rx}(\mathbf{r})\cdot \frac{\partial \mathbf{M}_0(\mathbf{r},t)}{\partial t} dV  ,
\end{aligned}
\end{equation}
where $\textbf{M}(\textbf{r}) = c(\textbf{r}) \textbf{M}_0(\textbf{r})$ and $c(\textbf{r})$ is concentration of SPION in the sample.

In the following, we assume perfect cancellation of the excitation magnetic field by the Rx gradiometer coil system (see Sec.~\ref{methods}), so that the signal comes only from the magnetization. Also, we consider a point source corresponding to a point-volume uniform sample of SPION $c(r)=N \delta (r-r_0)$, where $N$ is the number of particles in a volume $V_0$, so Eq.~\ref{rx_sig_eqn} becomes
\begin{equation}\label{emf}
  u(t) = -\mu_0 N\textbf{P}_{rx}\cdot \frac{\partial \mathbf{M}_0(\mathbf{r},t)}{\partial t}.
\end{equation}

The nonlinear response in the magnetization of the SPION is given by the Langevin model (see Fig.~\ref{fig:theory}):
\begin{equation}\label{field_eqn}
  \textbf{M}_0(t) = m\mathcal{L}(\beta |\textbf{H}(t)|)\hat{\textbf{H}},
\end{equation}
where $\mathcal{L}$ is the Langevin function, $\beta = \mu_0 m/k_{\rm B}T$ where $m$ is the saturation magnetic moment of the SPION, $T$ is the temperature, and $k_{\rm B}$ is Boltzmann's constant. 
Unlike a fluxgate magnetometer, MPM does not rely on magnetic saturation of the material for operation. Instead, optimal performance is achieved within the nonlinear portion of the Langevin curve (see Fig.~\ref{fig:theory}(a)).

The voltage measured in the receive coil is a periodic function with period $T=1/f$, where $f$ is the frequency of the excitation field. The harmonic content of the signal is characterized by the Fourier coefficients of Eq.~(\ref{rx_sig_eqn}), with the complex amplitude of the nth harmonic given by

\begin{equation}
    \label{real_for_eqn}
    \tilde{u}_n = 
    \int_0^{T} u(t)cos(2\pi nf t)\frac{dt}{T} - i\int_0^{T} u(t)sin(2\pi nf t)\frac{dt}{T}.
\end{equation}
In the absence of relaxation time effects of the SPION and phase delays caused by electronics, the voltage signal will be 90$^\circ$ out of phase with the applied excitation field. Consequently, the real part of the complex amplitude 
becomes zero, and the signal is entirely imaginary. This imaginary amplitude can 
take on both positive and negative values, depending on the applied magnetic 
fields and the sensitivity profile of the receive coil. 

The first harmonic has the highest amplitude; however, it is overlaid by a strong  
background from feedthrough of the excitation field. The second strongest 
amplitude is the third harmonic ($3f$) of the signal (see Fig.~\ref{fig:theory}(c)).

Assuming the magnetization is aligned along $\textbf{P}_{rx}$, for a small nonlinearity near $H_0=0$, we can rewrite Eq.~\ref{field_eqn} in one dimensional case as an expansion around $H(t)$ with $a_n$ expansion coefficients as
\begin{equation}\label{expand1}
M_0(H)=a_1 H(t) - a_3 H(t)^3 
= a_1 H(t) \left( 1 - \frac{a_3}{a_1} H(t)^2 \right) .
\end{equation}

On the other hand, for $H \ll H_s$ we have $M(H)= (\mu -1) H$, so we can define expansion coefficients in Eq.~\ref{expand1} as $a_1=(\mu-1)V_0$ and $a_3=(\mu-1)V_0/H_s^2$ to obtain
\begin{equation}\label{expand2}
M_0(H)=(\mu - 1) V_0 H [1 - (H/H_s)^2],
\end{equation}
where $\mu$ is permeability and $H_s$ saturation magnetic field. 

In the presence of an additional dc magnetic field $H(t)=H_{\rm ext}+H(t)$, where $H_{\rm ext}=B_{\rm ext}/\mu_0$ there will be even harmonics present in the signal due to the asymmetry of the magnetization response around zero of the M-H curve (see Fig.~\ref{fig:theory}(a-c)). Thus, keeping the second term in the expansion, for the second and third harmonics, we can get (see Appendix~\ref{app::harm_exp}):
\begin{equation}
    \label{expand3}
  \begin{aligned}
M_0^{(2)}=& 
\left[\frac{3}{2}(\mu-1)V_0 H_{\rm ext}\frac{H_0^2}{H_s^2} \right], \\
& M_0^{(3)}= 
\left[\frac{1}{4}(\mu-1) V_0 \frac{H_0^3}{H_s^2} \right].
\end{aligned}
\end{equation}
Further, we keep only time-dependent terms in $M_0 (t)$ Eq.~\ref{app:M0} that contribute to {\it emf}. Using Eqs.~\ref{emf}, \ref{expand2}, the amplitude of the second harmonic gives the dependence on the magnitude of the unknown magnetic field of interest as follows
\begin{equation}\label{emf_alt}
  u_2 = 6 \pi f N V_0 P_{rx} (\mu-1) \left(\frac{H_0}{H_s}\right)^2 B_{\rm ext} 
  = \alpha B_{\rm ext},
\end{equation}
with $\alpha$ - calibration sensitivity constant in units of ($\mu$V/$\mu$T).
Thus, by measuring the second-harmonic amplitude, we can obtain the external-field calibration curve. 

The third harmonic signal is similarly obtained from $M_0^{(3)}$ as
\begin{equation}\label{emf_u3}
  u_3 = \frac{3}{2} \pi f N V_0 P_{rx}(\mu-1)\left(\frac{H_0}{H_s}\right)^2B_0,
\end{equation}
where $B_0=\mu_0 H_0$. The normalized harmonic ratio is then
\begin{equation}\label{ratio}
  u_2 / u_3 = 4 B_{\rm ext}/B_0 .
\end{equation}
Therefore, the normalized calibration curve with known excitation field amplitude $B_0$ provides the external magnetic field $B_{\rm ext}$ independent of the SPION properties and the magnetometer's receive-coil profile.
%

To guide the design and interpretation of the magnetometer, we developed MATLAB-based simulations. The simulations model the magnetization response of the Langevin material and the resulting voltage induced in the receive coil. The simulations assume that the sensing medium obeys Langevin magnetization, as described by 
Eq.~\ref{field_eqn}, and that the relaxation to thermal equilibrium is instantaneous. Therefore, we do not model relaxation effects.

In this work, the simulations assume that the external dc field and the ac excitation field are aligned along the same direction. Under this assumption, the problem simplifies to a one-dimensional scalar expression $H(t)=H_{\mathrm{ext}}+H_0\cos(2\pi f t)$ and Eq.~\ref{field_eqn} becomes
\begin{equation}
    \label{eqn:mag_scalar}
    M_0(t)=m\mathcal{L}\left[\beta\left(H_{\mathrm{ext}}+H_0\cos(2\pi f t)\right)\right],
\end{equation}
where we use normalized fields $\beta H$.

The simulations further assume that the receive coil sensitivity is aligned with the applied magnetic fields. Therefore, Eq.~\ref{emf} simplifies to
\begin{equation}
    \label{eqn:urx_sim}
    u(t)=
    -\mu_0 N P_{rx}m \frac{d}{dt}\mathcal{L}\left[\beta\left(H_{\mathrm{ext}}+H_0\cos(2\pi f t)\right)\right].
\end{equation}
The amplitudes of the harmonic components for a given excitation amplitude $H_0$ and external field $H_{\mathrm{ext}}$ were then calculated numerically using a discrete version of Eq.~\ref{real_for_eqn}:
\begin{equation}
    \label{num_for_eqn}
    \tilde{u}_n=\frac{-i}{T}\sum_{t=0}^{T}u(t)\sin\left(2\pi n\frac{t}{T}\right).
\end{equation}
For each combination of $H_0$ and $H_{\mathrm{ext}}$, one excitation period was simulated using 100 time points. The amplitudes of the second and third harmonics were then calculated numerically. This procedure was repeated over a range of external dc fields and ac excitation amplitudes to generate the harmonic maps shown in Fig.~\ref{fig:theory}. As expected from the theory, the second harmonic is antisymmetric with respect to the external field and increases approximately linearly near zero field, while the third harmonic is symmetric and approximately constant for low dc fields. 

To estimate the magnetic fields, we set $\beta \approx 0.95~\mu_0/\mathrm{mT}$, approximately corresponding to the magnetization response of SPION (Synomag-D) \cite{Vogel2021} as used experimentally in this work. Then, the cross-section graph of second-harmonic amplitude vs ac excitation amplitude at a fixed dc offset field of $\sim$42 $\mu$T (Fig.~\ref{fig:theory}(f)) shows a maximum signal at $B_m$ = 4.5 mT.  
The third harmonic amplitude $u_3$, however, has monotonic growth with the excitation field and has to be taken into account in the ratio of harmonic amplitudes Eq.~\ref{ratio}.
\section{Methods}
\label{methods}
\begin{figure}[hbt!]
    \includegraphics[width=\linewidth]{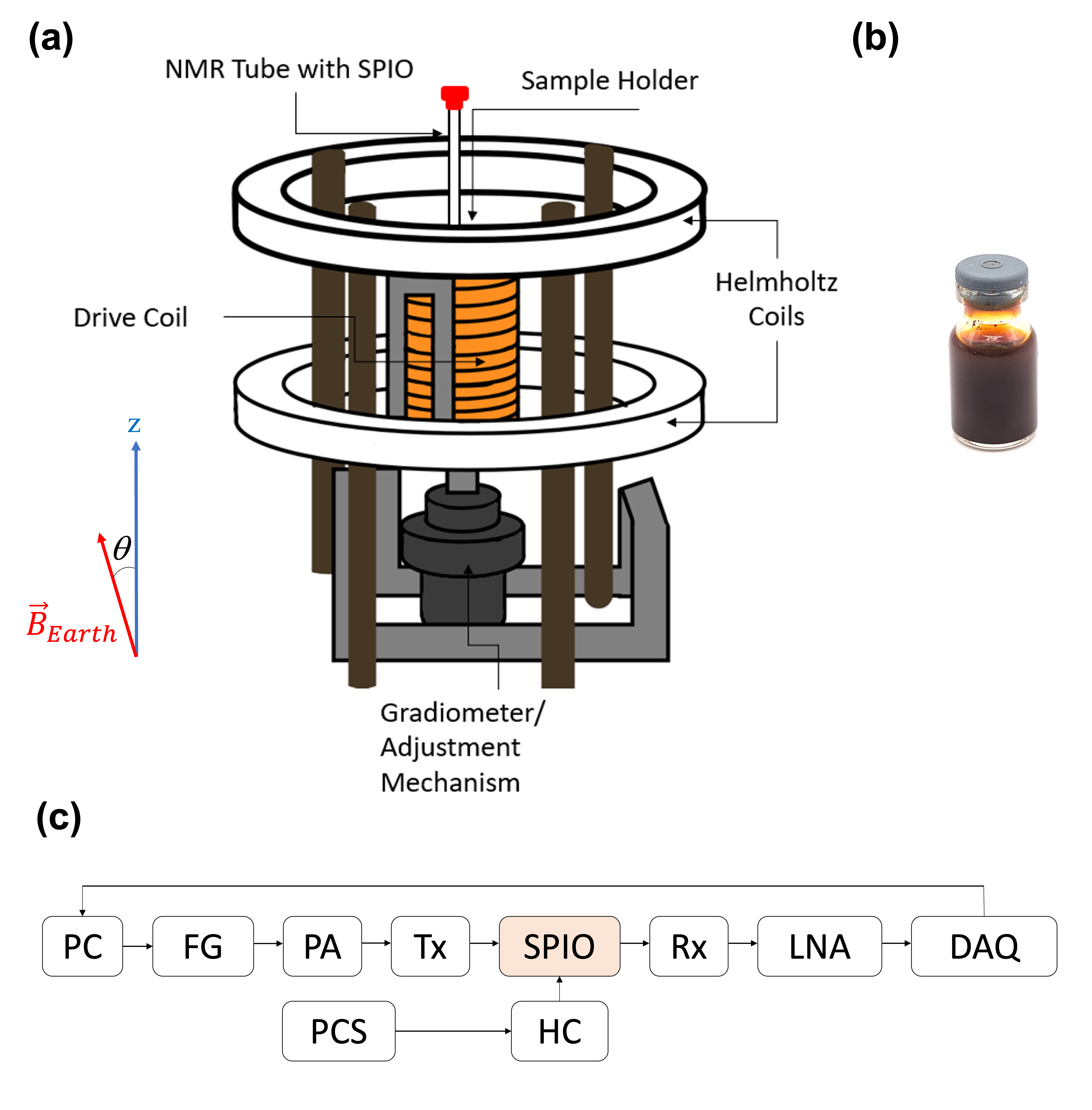}
    \caption{(a) Diagram of the MPM experimental setup with an NMR tube containing the SPIO sample and Helmholtz coils used for field calibration. (b) Example of SPION liquid sample in a vial. (c) Signal-chain block diagram: PC -- computer, FG -- function generator, PA -- power amplifier, Tx -- transmit coil, Rx -- receive coil, LNA -- low-noise amplifier, DAQ -- digital-analog converter, PCS -- precision current power supply, HC -- Helmholtz coils.}
    \label{setup}
\end{figure}
The proof-of-concept MPM instrumentation was constructed by integrating an MPS system \cite{Bastajian2024} with a pair of collinear Helmholtz coils, as shown in Fig.~\ref{setup}(a). 
The main components of the MPS are a drive/transmission (Tx) coil and a receive (Rx) coil (see details in Appendix~\ref{app:details}). The Tx solenoid coil generates a spatially uniform excitation magnetic field within the sample volume. A first-order solenoidal gradiometer serves as the Rx coil and allows manual tuning via a custom pulley-gear design. This configuration suppresses direct feedthrough from the excitation field while preserving the SPION sample signal. The SPION sample, in liquid suspension, is transferred from a vial into a 5-mm-diameter NMR tube. The tube is positioned within the sample holder so that the SPIO volume is centered relative to the Tx and Rx coils. 

In addition to MPS for MPM calibration, the Helmholtz coil pair was installed around the MPS and aligned coaxially with the Tx coil, with their geometric center coinciding with the sample position. A Precision Current Supply LDC202C (Thorlabs, NJ, USA) drove the coils. The current controller's bipolar output capability enabled measurements under both positive and negative applied magnetic fields. Calibration of the Helmholtz coils was performed using a Hall probe, yielding a magnetic field-to-current conversion factor (see Appendix \ref{app:details}).

All experiments were conducted using a 50~$\mu$L sample of PEG-coated Synomag-D SPIONs (Micromod GmbH, Rostock, Germany) with a hydrodynamic diameter of 70 nm and an iron concentration of 10 mg/mL (see Fig.~\ref{setup}(b)). 

The MPM signal chain is illustrated in Fig.~\ref{setup}(c). The system is controlled by a custom Python-based graphical user interface built with commonly available open-source packages, including CustomTkinter for the interface, PyVisa for instrument communication, NI-DAQmx for data acquisition, and NumPy for real-time signal processing. 
The sinusoidal signal is generated using a Keysight 33500B waveform generator (Keysight, Santa Rosa, CA, USA) at 4 kHz and amplified by a linear power amplifier, Techron 7226 (AE Techron, Elkhart, IN, USA). An impedance-matching circuit optimizes power transfer to the Tx coil. The ACS714 current sensor (Allegro MicroSystems, Manchester, NH, USA) detects the current through the Tx coil and converts it to field strength using pre-calibrated data. 

An excitation magnetic field induces a nonlinear magnetization response in the SPION sample, which is detected through its harmonic components. The voltage induced in the Rx coil is amplified with a low-noise voltage preamplifier, the SR560 (Stanford Research Systems, Sunnyvale, CA, USA), and then digitized by the NI USB-6366 DAQ card (National Instruments, Austin, Texas, USA). The data are recorded over 2000 periods at a sampling rate of 100 kHz. Fast Fourier Transform (FFT) algorithm converts the signal to the frequency domain and enables real-time processing using Python-based software. Additionally, the background signal without SPIO was recorded and digitally subtracted from each harmonic signal to eliminate any nonlinear components generated by the signal chain.

The harmonic response of the SPION sample was characterized as a function of an externally applied dc magnetic field. 
The dc field was generated by a Helmholtz coil pair. The applied current was incrementally varied between $\pm 200$ mA, corresponding to an external magnetic field range of $\pm 500~\mu$T.
The resulting calibration curve was used to evaluate the performance of the MPM. Instrument sensitivity, stability, and accuracy were assessed by continuous measurement over an extended period of time while recording the normalized harmonic response. 

The measured magnetic field values were converted using the calibration ratio and compared with readings from available commercial magnetometers (see Appendix~\ref{app:add_data}). This comparison enabled assessment of the MPM's accuracy and precision within the range of the Earth's geomagnetic field. In addition, vector measurements acquired using Hall probes were used to determine the orientation ($\theta$ in Fig.~\ref{setup}(a)) of Earth's magnetic field.

\section{Experimental results and discussions} 
\label{results}
\begin{figure*}[htb!]
    \includegraphics[width=\linewidth]{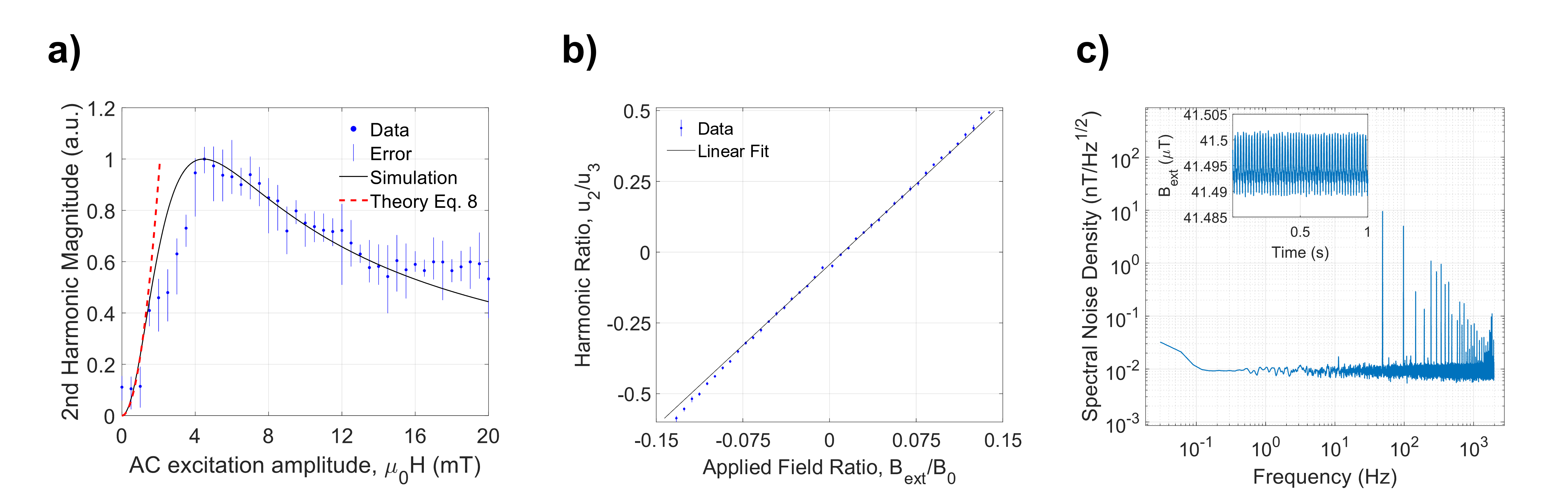}
    \caption{(a) Experimental data showing second harmonic amplitude vs ac excitation magnetic field; curve fit is a simulation from Fig.~\ref{fig:theory}(f). (b) Magnetic field calibration curve using harmonic ratio vs applied field ratio. The linear fit applies to the range of $\pm$200 $\mu$T with $R^2$ = 0.9984. (c) Spectral noise density (SND) of MPM. The inset shows an interval of timetrace of the measured external magnetic field. }
    \label{mpmdata}
\end{figure*}
To determine the optimal excitation conditions for magnetometry, a preliminary study was conducted to characterize the dependence of the second-harmonic response on the amplitude of the ac excitation field while keeping the dc offset field constant. 
The ac excitation field, $B_0$, was varied from 0 to 20 mT in increments of 0.5 mT, and the measurement sequence was repeated over five independent runs. The second harmonic amplitude was found to reach a maximum at an ac excitation field of 4.5 mT, as shown in Fig.~\ref{mpmdata}(a). 
The simulation curve was fit to the data using $\beta$  as a free parameter and the theoretical curve of Eq.~\ref{emf_alt} with $\nu$ as a fitting parameter according to $u_2=\nu (\beta H_0)^2$ to coincide with the simulation at low $H_0$. The fits yielded $\beta \approx$ 1 $\mu_0$/mT and $\nu$ = 0.2311 $\mu$V.

Following optimization of the excitation conditions, the response of the SPION sample was characterized as a function of an externally applied dc magnetic field with $\mu_0 H_0$ = 3.5651 mT. The amplitudes of the second and third harmonics were extracted from the measured spectra and plotted versus the ratio of the applied dc magnetic field and the ac drive field amplitude to obtain a calibration curve for MPM. Figure~\ref{mpmdata}(b) shows the experimentally measured normalized harmonic response within the extended dc operating regime of $\pm$ 500 $\mu$T. A linear fit to the data over the $\pm$ 200 $\mu$T linear operating range follows the equation:
\begin{equation}
    \label{norm_calibration}
    {u_2}/{u_3} = 3.7978\times {B_{\rm ext}}/{B_0}-0.04420 ,
\end{equation}
where ${u_2}/{u_3}$ is the normalized harmonic ratio and $B_{ext}/B_0$ is the normalized applied field. The difference between the obtained slope and the theoretical value of 4 can be attributed to the deviation of experimental conditions from the theoretical approximation of Eq.~\ref{ratio}. 
From the linear fit, the zero crossing occurred at 41.492~$\mu$T, corresponding to the local vertical component of the Earth's geomagnetic field at the SPION location.

With the acquired time-domain signal over 10 min, the spectral noise density (SND) of the MPM was plotted as shown in Fig.~\ref{mpmdata}(c). 
Apart from the expected $1/f$ noise at low frequencies, the SND remains constant at 10.8 pT/$\sqrt{\text{Hz}}$ across the frequency range, with a noise density of 10.4 pT/$\sqrt{\text{Hz}}$ at 1 Hz.  
Thus, the estimated sensitivity of MPM is comparable to that of other high-end magnetometers, as shown in Fig.~\ref{compare}. 
\begin{table}[htb!]
\begin{tabular}{c c}
\hline
Magnetometer & $B_z$ Measurement ($\mu$T)\\
\hline
SENIS 3MTS (Hall Probe) & 43.6 $\pm$ 1.3\\
HGM09s Gaussmeter (Hall Sensor) & 41.40 $\pm$ 0.08\\
Twinleaf OMG (OPM)*    & 41.396 $\pm$ 0.002\\
MPM    & 41.495 $\pm$ 0.002\\
\hline
\end{tabular}
\caption{Comparison of various magnetometer types for measurements of the z-component of Earth's magnetic field. The $B_z$ measurement is obtained by averaging the trace in Fig.~\ref{stablecomp} over a 10 min acquisition time. *The OPM $B_z$ component is estimated from the total magnetic field by obtaining the angle $\theta$ of the Earth's magnetic field using Hall probe magnetometers to measure two orthogonal components (see Appendix~\ref{app:add_data}).
}
\label{zcomp}
\end{table}
In addition, we compared the measured z-component of Earth's magnetic field with measurements from commercial magnetometers to evaluate the MPM's accuracy, as summarized in Table~\ref{zcomp}. For each instrument, the reported magnetic field is the mean value measured over the 10 min trace acquisition time, with the uncertainty given by one standard deviation.
Several magnetometers were used according to the procedure in Appendix \ref{app:add_data}.
Assuming the OPM provides the highest accuracy, a 0.2\% error can be explained by imperfections in the orientation of the MPS coils relative to the vertical and uncertainty in the probe location.

The proof-of-concept measurements validate the theoretical framework of magnetic particle magnetometry and demonstrate direct detection of the vertical component of the Earth's magnetic field. In its current configuration, the MPM measures the magnetic field along a single axis defined by the orientation of the excitation and detection coils. Vector 3D magnetic field measurements could be achieved by rotating the sensor assembly or by employing multiple sets of orthogonally oriented coils to independently measure the field components $B_x$, $B_y$, and $B_z$.

The technique's sensitivity depends strongly on the magnetic properties of the nanoparticle tracer.
Nanoparticles with higher saturation magnetization, including composite heterometallic nanoparticles beyond conventional iron oxides, may provide enhanced sensitivity in applications where biocompatibility is not a primary constraint \cite{Ferguson2009, Hart2026}. In addition, nanoparticle relaxation processes, which depend on particle size, anisotropy, and hydrodynamic environment \cite{Brown1963, COFFEY1994, Reeves2015}, may ultimately limit the magnetometer's temporal response.

The MPM architecture is inherently scalable and can be adapted to a wide range of sensing geometries. For compact sensing applications, the excitation and detection coils may be miniaturized to reduce the overall system footprint. Alternatively, enlarged excitation and receive coils can enable remote magnetic-field measurements over larger volumes. In such configurations, the nanoparticle sample can serve as a passive magnetic-field transducer within the volume of interest, while the interrogation hardware remains spatially separated from the measurement location. Furthermore, nanoparticles may be embedded directly within an object or structure, enabling localized magnetic field measurements with either volumetric solenoidal coils or surface-coil geometries \cite{Tonyushkin2017a, McDonough2022b}.
\section{Conclusions}
In this work, we demonstrated a magnetic particle magnetometer (MPM) that exploits the nonlinear Langevin magnetization response of superparamagnetic iron oxide nanoparticles (SPIONs). The proposed sensor utilizes the even-order harmonic components of the nanoparticle magnetization, particularly the second harmonic, as a spectral signature directly related to the local magnetic field. By remotely detecting these harmonic responses, the technique enables localized magnetic field measurements without requiring electrical connections to the sensing region.
The proof-of-concept system offers a linear operating range of $\pm 200~\mu$T and achieves a high sensitivity of 10~pT/$\sqrt{\mathrm{Hz}}$ in the unshielded environment. Using the MPM, we measured the local vertical component of the Earth's magnetic field and obtained results consistent with those from commercially available magnetometers, including optically pumped magnetometers (OPMs).
Unlike conventional magnetometers, the proposed approach combines the robustness and spectral specificity of fluxgate-based detection with the embedded and remote-sensing capabilities typical of quantum sensors, while avoiding cryogenic cooling or complex optical instrumentation. Since SPIO nanoparticles are biocompatible and can be distributed throughout a medium, the technique enables high-resolution sensing in biological tissues, inaccessible structures, and electromagnetically noisy environments through carrier-frequency-selective harmonic detection. 
\begin{acknowledgments}
This work was supported in part by the NIH under Grant R15EB028535.
\end{acknowledgments}
\appendix
%
\section{Derivation of the Harmonic Expansion}
\label{app::harm_exp}
Starting from the cubic approximation of the Langevin response,
\begin{equation}
M_0(H)=a_1H-a_3H^3 ,
\end{equation}
and assuming that the applied magnetic field is the sum of a dc offset field and an ac excitation field,
\begin{equation}
H(t)=H_{\rm ext}+H_0\sin(\omega t),
\end{equation}
where $\omega=2\pi f$, the magnetization can be written as
\begin{equation}
M_0(t)=a_1\left(H_{\rm ext}+H_0\sin\omega t\right)
-a_3\left(H_{\rm ext}+H_0\sin\omega t\right)^3 .
\end{equation}

Expanding the cubic term gives
\begin{equation}
\begin{aligned}
M_0(t)=&\, a_1H_{\rm ext}+a_1H_0\sin\omega t
-a_3H_{\rm ext}^3
-3a_3H_{\rm ext}^2H_0\sin\omega t \\
&-3a_3H_{\rm ext}H_0^2\sin^2\omega t
-a_3H_0^3\sin^3\omega t .
\end{aligned}
\end{equation}

Using the identities
\begin{equation}
\sin^2\omega t=\frac{1-\cos(2\omega t)}{2},
\qquad
\sin^3\omega t=\frac{3\sin(\omega t)-\sin(3\omega t)}{4},
\end{equation}
we obtain
\begin{equation}
\begin{aligned}
M_0(t)=&\, a_1H_{\rm ext}+a_1H_0\sin\omega t
-a_3H_{\rm ext}^3
-3a_3H_{\rm ext}^2H_0\sin\omega t \\
&-\frac{3}{2}a_3H_{\rm ext}H_0^2
+\frac{3}{2}a_3H_{\rm ext}H_0^2\cos(2\omega t) \\
&-\frac{3}{4}a_3H_0^3\sin(\omega t)
+\frac{1}{4}a_3H_0^3\sin(3\omega t).
\end{aligned}
\end{equation}

Collecting terms with the same frequency gives
\begin{equation}
\begin{aligned}
M_0(t)=&
\left[a_1H_{\rm ext}-a_3H_{\rm ext}^3
-\frac{3}{2}a_3H_{\rm ext}H_0^2\right] \\
&+\left[a_1H_0-3a_3H_{\rm ext}^2H_0
-\frac{3}{4}a_3H_0^3\right]\sin(\omega t) \\
&+\frac{3}{2}a_3H_{\rm ext}H_0^2\cos(2\omega t)
+\frac{1}{4}a_3H_0^3\sin(3\omega t).
\end{aligned}
\end{equation}

Finally, substituting $\omega=2\pi f$ yields
\begin{equation}\label{app:M0}
\begin{aligned}
M_0(t)=&
\left[a_1H_{\rm ext}-a_3H_{\rm ext}^3
-\frac{3}{2}a_3H_{\rm ext}H_0^2\right] \\
&+\left[a_1H_0-3a_3H_{\rm ext}^2H_0
-\frac{3}{4}a_3H_0^3\right]\sin(2\pi ft) \\
&+\frac{3}{2}a_3H_{\rm ext}H_0^2\cos(2\pi 2ft)
+\frac{1}{4}a_3H_0^3\sin(2\pi 3ft).
\end{aligned}
\end{equation}
%
\section{Additional Details on MPS}
\label{app:details}
Here, we provide additional details on the coils composing the MPS apparatus.
The transmit (Tx) coil consists of a 136 mm-long acrylic tube with an inner diameter of 63.5 mm, wound with six layers of 106 turns of 41/36 continuous Litz wire. The receive (Rx) coil is wound on an 18 mm-long 3D-printed nylon former using 20 turns of 160/44 Litz wire. The Rx coil was designed to minimize the separation between the coil and the SPION sample, thereby maximizing detection sensitivity while reducing direct coupling (feedthrough) from the excitation field. A cancellation coil comprising a first-order gradiometer, fabricated from 3D-printed nylon, is positioned coaxially with both the Tx and Rx coils. The gradiometer winding is connected in series with the Rx coil with reversed polarity, enabling cancellation of the excitation field feedthrough while preserving the magnetic response generated by the SPION sample. The gradiometer position is mechanically adjustable along the coil axis, allowing fine-tuning of feedthrough cancellation to maximize common-mode rejection.
\begin{figure}[htb!]
    \includegraphics[width=0.85\linewidth]{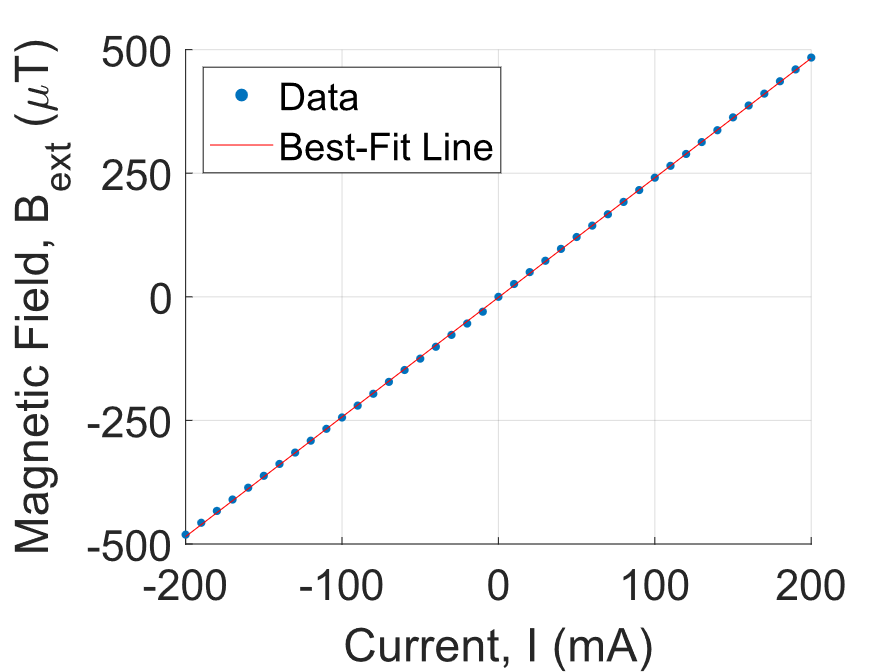}
    \caption{Calibration curve of Helmholtz Coils using HGM09s gaussmeter. 
    Linear fit is given by the equation $B_{\rm ext} = 2.419 \times I-1.267$, with $R^2$ = 0.9999.}
    \label{HHcoil}
\end{figure}
%
The Helmholtz coil pair consists of two coils, each wound with 609 turns, having an inner diameter of 270 mm and a center-to-center spacing of 120 mm. The coil assembly is mounted on an adjustable support, allowing alignment with the SPION sample in both the horizontal and vertical directions. This alignment ensures that the sample is located at the center of the Helmholtz coils, where the magnetic field is most uniform.
The Helmholtz coils were calibrated using an HGM09s gaussmeter (MAGSYS GmbH, Dortmund, Germany). The measured relationship between the applied current and the magnetic field at the center of the coil pair is (see Fig.~\ref{HHcoil}) $B_{\rm ext} = 2.419 \times I-1.267 , $
where $B_{\rm ext}$ is the magnetic field at the center of the Helmholtz coils, and {\em I} is the current supplied to the coils.
%
%
\section{Additional Data on Stability of MPM}
\label{app:add_data}
\begin{figure}[htb!]
    \includegraphics[width=0.75\linewidth]{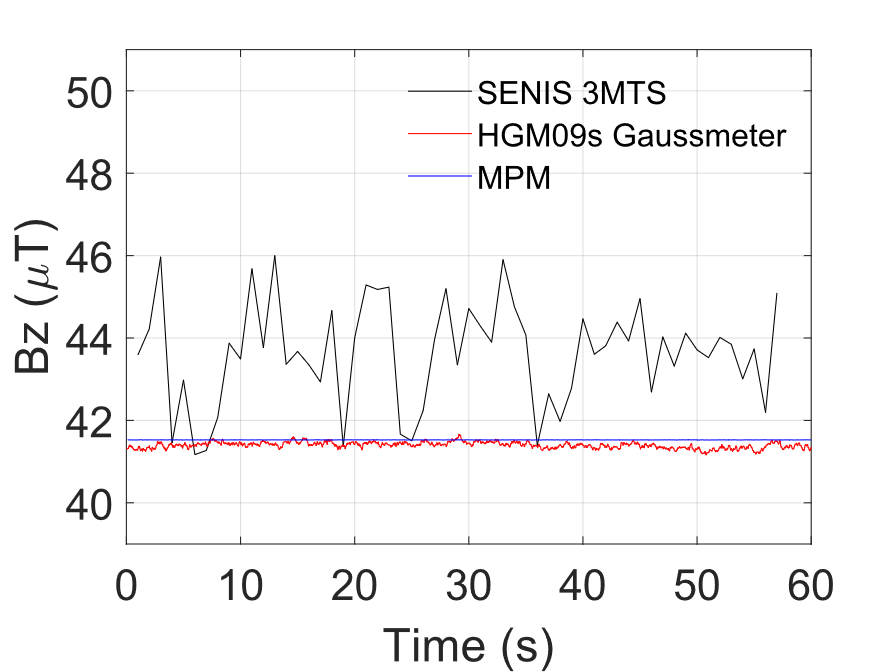}
    \caption{Comparison of Hall probe magnetometers and MPM: measurement accuracy and stability over 60 s.}
    \label{stablecomp}
\end{figure}
In this section, we provide additional data on the accuracy and stability of MPM compared with other commercially available magnetometers. Reference measurements were acquired using a SENIS 3MTS Hall probe (GMW Associates, San Carlos, CA, USA), an HGM09s gaussmeter (MAGSYS GmbH, Dortmund, Germany), and a Twinleaf OMG (Twinleaf LLC, Plainsboro, NJ, USA) OPM. Since the OPM  is the total field magnetometer, $B_{\rm tot}$, vector measurements from the Hall probes were used to determine the orientation of Earth's magnetic field and to calculate the corresponding $B_z$ component. All measurements were performed within a common laboratory coordinate system, with the sensor axes aligned to measure the same magnetic field components. The laboratory's geographical location is given by the following GPS coordinates: 42.67090$^\circ$ N, 83.21790$^\circ$ W; however, the measurement is strongly impacted by the surroundings of an unshielded room. 

The SENIS 3MTS was operated using the manufacturer's ``Start 3MTS'' software with the measurement range set to the lowest available range of 100 mT. Data were acquired at a sampling rate of 10 Sa/s. The probe was oriented to measure the z-component of Earth's magnetic field, $B_z$. Since the 3MTS simultaneously measures $B_x$, $B_y$, and $B_z$, it provides a direct vector measurement of the ambient geomagnetic field and was used to determine the field orientation required for comparison with scalar magnetometer measurements. Mean values of $B_x$ and $B_z$ were calculated over a 60 s acquisition interval to determine the angle between the geomagnetic field vector and the z-axis, $\theta$ = 26.853$^\circ$.

The HGM09s single-axis gaussmeter was operated using a Python acquisition program that communicated with the instrument through its serial interface. Data were collected via USB at a sampling rate of 48 Sa/s with the measurement range set to the lowest available range of 10 mT. The probe was initially aligned to measure $B_z$ and subsequently rotated by 90$^\circ$ to measure the horizontal field component, $B_x$. These measurements independently verified the field components measured by the 3MTS and provided an additional reference for comparison with the MPM measurements.

The Twinleaf OMG has gradiometer configuration with two independent magnetometers separated by 1 cm base. Both magnetometers recorded the total field magnitudes, which were averaged via a LabVIEW interface (National Instruments, Austin, Texas, USA). 
The average value $B_{\rm tot}$ = 44.929 $\mu$T measured by the OMG over the same interval was then projected onto the z-axis to obtain the corresponding magnetic field component, which is reported in Table~\ref{zcomp}. This approach allowed direct comparison of the OPM measurements with the z-component measured by the MPM.
%
%
\bibliographystyle{apsrev4-2}
\bibliography{mpm}

@PREAMBLE{
 "\providecommand{\noopsort}[1]{}" 
 # "\providecommand{\singleletter}[1]{#1}%" 
}

@book{clarke2004,
  title={The SQUID Handbook: Fundamentals and technology of SQUIDs and SQUID systems},
  author={Clarke, J. and Braginski, A.I.},
  number={v. 1},
  url={https://books.google.com/books?id=qXcBzQEACAAJ},
  year={2004},
  publisher={Wiley-VCH}
}

@article{Sternickel2006,
doi = {10.1088/0953-2048/19/3/024},
url = {https://doi.org/10.1088/0953-2048/19/3/024},
year = {2006},
month = {feb},
publisher = {},
volume = {19},
number = {3},
pages = {S160},
author = {Sternickel, Karsten and Braginski, Alex I},
title = {Biomagnetism using SQUIDs: status and perspectives},
journal = {Superconductor Science and Technology}
}

@ARTICLE{Lenz2006,
  author={Lenz, J. and Edelstein, S.},
  journal={IEEE Sensors Journal}, 
  title={Magnetic sensors and their applications}, 
  year={2006},
  volume={6},
  number={3},
  pages={631-649},
  doi={10.1109/JSEN.2006.874493}
}

@article{MAPPS1997,
title = {Magnetoresistive sensors},
journal = {Sensors and Actuators A: Physical},
volume = {59},
number = {1},
pages = {9-19},
year = {1997},
note = {1st European magnetic sensors and actuators conference},
issn = {0924-4247},
doi = {https://doi.org/10.1016/S0924-4247(97)80142-2},
url = {https://www.sciencedirect.com/science/article/pii/S0924424797801422},
author = {D.J. Mapps}
}

@ARTICLE{Zheng2019,
  author={Zheng, Chao and Zhu, Ke and Cardoso de Freitas, Susana and Chang, Jen-Yuan and Davies, Joseph E. and Eames, Peter and Freitas, Paulo P. and Kazakova, Olga and Kim, CheolGi and Leung, Chi-Wah and Liou, Sy-Hwang and Ognev, Alexey and Piramanayagam, S. N. and Ripka, Pavel and Samardak, Alexander and Shin, Kwang-Ho and Tong, Shi-Yuan and Tung, Mean-Jue and Wang, Shan X. and Xue, Songsheng and Yin, Xiaolu and Pong, Philip W. T.},
  journal={IEEE Transactions on Magnetics}, 
  title={Magnetoresistive Sensor Development Roadmap (Non-Recording Applications)}, 
  year={2019},
  volume={55},
  number={4},
  pages={1-30},
  doi={10.1109/TMAG.2019.2896036}}

@article{RIPKA1992,
title = {Review of fluxgate sensors},
journal = {Sensors and Actuators A: Physical},
volume = {33},
number = {3},
pages = {129-141},
year = {1992},
issn = {0924-4247},
doi = {https://doi.org/10.1016/0924-4247(92)80159-Z},
url = {https://www.sciencedirect.com/science/article/pii/092442479280159Z},
author = {Pavel Ripka}
}

@Inbook{Janosek2017,
author="Janosek, Michal",
editor="Grosz, Asaf
and Haji-Sheikh, Michael J.
and Mukhopadhyay, Subhas C.",
title="Parallel Fluxgate Magnetometers",
bookTitle="High Sensitivity Magnetometers",
year="2017",
publisher="Springer International Publishing",
address="Cham",
pages="41--61",
isbn="978-3-319-34070-8",
doi="10.1007/978-3-319-34070-8_2",
url="https://doi.org/10.1007/978-3-319-34070-8_2"
}

@article{Jensen2016,
author = {Jensen, Kasper and Budvytyte, Rima and Thomas, Rodrigo A and Wang, Tian and Fuchs, Annette M and Balabas, Mikhail V and Vasilakis, Georgios and Mosgaard, Lars D and St{\ae}rkind, Hans C and Muller, Jorg H. and Heimburg, Thomas and Olesen, S{\o}ren Peter and Polzik, Eugene S.},
doi = {10.1038/srep29638},
issn = {20452322},
journal = {Scientific Reports},
number = {July},
pages = {1--7},
publisher = {Nature Publishing Group},
title = {{Non-invasive detection of animal nerve impulses with an atomic magnetometer operating near quantum limited sensitivity}},
volume = {6},
year = {2016}
}

@article{Budker2007,
   author = {Dimtry. Dmitry Budker and Michael. Romalis},
   doi = {10.1038/nphys566},
   isbn = {1745-2473},
   issn = {1745-2473},
   issue = {4},
   journal = {Nature Physics},
   pages = {227-234},
   pmid = {15003161},
   title = {Optical magnetometry},
   volume = {3},
   url = {http://dx.doi.org/10.1038/nphys566},
   year = {2007}
}

@ARTICLE{Colombo2020,
  author={S. {Colombo} and V. {Lebedev} and A. {Tonyushkin} and S. {Pengue} and A. {Weis}},
  journal={IEEE Transactions on Medical Imaging}, 
 title={Imaging Magnetic Nanoparticle Distributions by Atomic Magnetometry-Based Susceptometry}, 
  year={2020},
  volume={39},
  number={4},
  pages={922-933},
  doi={10.1109/TMI.2019.2937670},
  ISSN={1558-254X},
  month={April}}

@article{Colombo2016a,
author = {Colombo, Simone and Lebedev, Victor and Grujic, Zoran D. and Dolgovskiy, Vladimir and Weis, Antoine},
doi = {10.18416/ijmpi.2016.1604001},
journal = {International Journal on Magnetic Particle Imaging},
number = {1},
pages = {1604001},
title = {{M(H) dependence and size distribution of SPIONs measured by atomic magnetometry}},
volume = {2},
year = {2016}
}

@article{Xu2006,
author = {Xu, S and Donaldson, M.H. and Pines, A and Rochester, S.M. and Budker, D. and Yashchuk, V.V.},
doi = {10.1063/1.2400077},
issn = {00036951},
journal = {Applied Physics Letters},
number = {22},
pages = {1--10},
title = {{Application of atomic magnetometry in magnetic particle detection}},
volume = {89},
year = {2006}
}

@article{Limes2020,
  title = {Portable Magnetometry for Detection of Biomagnetism in Ambient Environments},
  author = {Limes, M.E. and Foley, E.L. and Kornack, T.W. and Caliga, S. and McBride, S. and Braun, A. and Lee, W. and Lucivero, V.G. and Romalis, M.V.},
  journal = {Phys. Rev. Appl.},
  volume = {14},
  issue = {1},
  pages = {011002},
  numpages = {6},
  year = {2020},
  month = {Jul},
  publisher = {American Physical Society},
  doi = {10.1103/PhysRevApplied.14.011002},
  url = {https://link.aps.org/doi/10.1103/PhysRevApplied.14.011002}
}

@article{Gleich2005,
  title={Tomographic imaging using the nonlinear response of magnetic particles},
  author={Gleich, Bernhard and Weizenecker, J{\"u}rgen},
  journal={Nature},
  volume={435},
  number={7046},
  pages={1214},
  year={2005},
  doi = {10.1038/nature03808},
  url = {https://doi.org/10.1038/nature03808},
  publisher={Nature Publishing Group}
}

@article{Panagiotopoulos2015,
author = {Panagiotopoulos, Nikolaos and Vogt, Florian and Barkhausen, J{\"{o}}rg and Buzug, Thorsten M and Duschka, Robert L and L{\"{u}}dtke-Buzug, Kerstin and Ahlborg, Mandy and Bringout, Gael and Debbeler, Christina and Gr{\"{a}}ser, Matthias and Kaethner, Christian and Stelzner, Jan and Medimagh, Hanne and Haegele, Julian},
doi = {10.2147/IJN.S70488},
issn = {1178-2013},
journal = {International Journal of Nanomedicine},
month = {apr},
pages = {3097},
pmid = {25960650},
title = {{Magnetic particle imaging: current developments and future directions}},
year = {2015}
}

@article{McDonough2022b,
		Author = {McDonough, C and Pagan, J and Tonyushkin, A},
	Journal = {Physics in Medicine \& Biology},
	Month = {dec},
	Number = {24},
	Pages = {245009},
	Publisher = {IOP Publishing},
	Title = {Implementation of the surface gradiometer receive coils for the improved detection limit and sensitivity in the single-sided MPI scanner},
	volume = {67},
	year = {2022},
	doi = {10.1088/1361-6560/aca5ec}
 }

@inproceedings{Bastajian2024,
   author = {Bastajian, C. and McDonough, C and Chrisekos, J. and Tonyushkin, A.},
   booktitle = {Bulletin of the American Physical Society},
   month = {4},
   title = {Magnetic Particle Spectrometer for Characterization of Superparamagnetic Iron Oxide Nanoparticles},
   url = {https://meetings-archive.aps.org/eglss/2024/q03/2/},
   year = {2024}
}

@article{Lowa2017,
author = {L{\"{o}}wa, Norbert and Seidel, Maria and Radon, Patricia and Wiekhorst, Frank},
doi = {10.1016/j.jmmm.2016.10.096},
issn = {03048853},
journal = {Journal of Magnetism and Magnetic Materials},
number = {June 2016},
pages = {133--138},
publisher = {Elsevier},
title = {{Magnetic nanoparticles in different biological environments analyzed by magnetic particle spectroscopy}},
volume = {427},
year = {2017}
}

@article{Barry2016,
author = {John F. Barry  and Matthew J. Turner  and Jennifer M. Schloss  and David R. Glenn  and Yuyu Song  and Mikhail D. Lukin  and Hongkun Park  and Ronald L. Walsworth },
title = {Optical magnetic detection of single-neuron action potentials using quantum defects in diamond},
journal = {Proceedings of the National Academy of Sciences},
volume = {113},
number = {49},
pages = {14133-14138},
year = {2016},
doi = {10.1073/pnas.1601513113},
URL = {https://www.pnas.org/doi/abs/10.1073/pnas.1601513113},
eprint = {https://www.pnas.org/doi/pdf/10.1073/pnas.1601513113}
}

@article{Taylor2008,
   author = {J M Taylor and P Cappellaro and L Childress and L Jiang and D Budker and P R Hemmer and A Yacoby and R Walsworth and M D Lukin},
   doi = {10.1038/nphys1075},
   issn = {1745-2481},
   issue = {10},
   journal = {Nature Physics},
   pages = {810-816},
   title = {High-sensitivity diamond magnetometer with nanoscale resolution},
   volume = {4},
   url = {https://doi.org/10.1038/nphys1075},
   year = {2008}
}

@article{Elmore1938,
   author = {W. C. Elmore},
   doi = {10.1103/PhysRev.54.1092},
   issn = {0031-899X},
   issue = {12},
   journal = {Physical Review},
   month = {12},
   pages = {1092-1095},
   title = {The Magnetization of Ferromagnetic Colloids},
   volume = {54},
   year = {1938}
}

@article{Kolosnjaj-Tabi2016a,
author = {Kolosnjaj-Tabi, Jelena and Lartigue, L??naic and Javed, Yasir and Luciani, Nathalie and Pellegrino, Teresa and Wilhelm, Claire and Alloyeau, Damien and Gazeau, Florence},
doi = {10.1016/j.nantod.2015.10.001},
isbn = {17480132},
issn = {1878044X},
journal = {Nano Today},
number = {3},
pages = {280--284},
publisher = {Elsevier Ltd},
title = {{Biotransformations of magnetic nanoparticles in the body}},
volume = {11},
year = {2016}
}

@article{Stabi2011,
author = {Stabi, Katie L and Bendz, Lisa M},
doi = {10.1345/aph.1q431},
issn = {1060-0280},
journal = {Annals of Pharmacotherapy},
month = {dec},
number = {12},
pages = {1571--1575},
pmid = {22045905},
title = {{Ferumoxytol Use as an Intravenous Contrast Agent for Magnetic Resonance Angiography}},
volume = {45},
year = {2011}
}

@article{Ferguson2009,
author = {Ferguson, R. Matthew and Minard, Kevin R. and Krishnan, Kannan M.},
doi = {10.1016/j.jmmm.2009.02.083},
issn = {03048853},
journal = {Journal of Magnetism and Magnetic Materials},
month = {may},
number = {10},
pages = {1548--1551},
pmid = {23106153},
title = {{Optimization of nanoparticle core size for magnetic particle imaging}},
volume = {321},
year = {2009}
}

@article{Reeves2015,
author = {Reeves, Daniel B. and Weaver, John B.},
doi = {10.1063/1.4936930},
issn = {00036951},
journal = {Applied Physics Letters},
number = {22},
title = {{Combined Neel and Brown rotational Langevin dynamics in magnetic particle imaging, sensing, and therapy}},
volume = {107},
year = {2015}
}

@article{Dhavalikar2015,
author = {Dhavalikar, Rohan and Maldonado-Camargo, Lorena and Garraud, Nicolas and Rinaldi, Carlos},
doi = {10.1063/1.4935158},
issn = {10897550},
journal = {Journal of Applied Physics},
number = {17},
pmid = {26576063},
title = {{Ferrohydrodynamic modeling of magnetic nanoparticle harmonic spectra for magnetic particle imaging}},
volume = {118},
year = {2015}
}

@article{Brown1963,
  title = {Thermal Fluctuations of a Single-Domain Particle},
  author = {Brown, William Fuller},
  journal = {Phys. Rev.},
  volume = {130},
  issue = {5},
  pages = {1677--1686},
  numpages = {0},
  year = {1963},
  month = {Jun},
  publisher = {American Physical Society},
  doi = {10.1103/PhysRev.130.1677},
  url = {https://link.aps.org/doi/10.1103/PhysRev.130.1677}
}

@article{COFFEY1994,
title = {Simple approximate formulae for the magnetic relaxation time of single domain ferromagnetic particles with uniaxial anisotropy},
journal = {Journal of Magnetism and Magnetic Materials},
volume = {131},
number = {3},
pages = {L301-L303},
year = {1994},
issn = {0304-8853},
doi = {https://doi.org/10.1016/0304-8853(94)90272-0},
url = {https://www.sciencedirect.com/science/article/pii/0304885394902720},
author = {W.T. Coffey and P.J. Cregg and D.S.F. Crothers and J.T. Waldron and A.W. Wickstead}
}

@article{Garcia1998,
  title = {Langevin-dynamics study of the dynamical properties of small magnetic particles},
  author = {Garc\'{\i}a-Palacios, Jos\'e Luis and L\'azaro, Francisco J.},
  journal = {Phys. Rev. B},
  volume = {58},
  issue = {22},
  pages = {14937--14958},
  numpages = {0},
  year = {1998},
  month = {Dec},
  publisher = {American Physical Society},
  doi = {10.1103/PhysRevB.58.14937},
  url = {https://link.aps.org/doi/10.1103/PhysRevB.58.14937}
}

@article{Hart2026,
doi = {10.1088/2057-1976/ae44a0},
url = {https://doi.org/10.1088/2057-1976/ae44a0},
year = {2026},
month = {feb},
publisher = {IOP Publishing},
volume = {12},
number = {2},
pages = {022001},
author = {Hart, Jaiden and Nguyen T Tran, Linh and Ena, Tamara Faranaz and Natekar, Niranjan A and Rezaei, Bahareh and Jiao, Yipeng and Zuo, Hansong and Wang, Hanlei and Chugh, Vinit and Azizi, Ebrahim and Karampelas, Ioannis H and He, Rui and Gomez-Pastora, Jenifer and Wu, Kai},
title = {Magnetic nanoparticles for cancer theranostics},
journal = {Biomedical Physics \& Engineering Express}
}

@article{Vogel2021,
   author = {P. Vogel and T. Kampf and M. A. Ruckert and C. Grüttner and A. Kowalski and H. Teller and V. C. Behr},
   doi = {10.18416/IJMPI.2021.2103003},
   issn = {23659033},
   issue = {1},
   journal = {International Journal on Magnetic Particle Imaging},
   publisher = {Infinite Science Publishing},
   title = {Synomag: The new high-performance tracer for magnetic particle imaging},
   volume = {7},
   year = {2021}
}

@article{Tonyushkin2017a,
    title = {{Single-Sided Field-Free Line Generator Magnet for Multi-Dimensional Magnetic Particle Imaging}},
    year = {2017},
    journal = {IEEE Transactions on Magnetics},
    author = {Tonyushkin, Alexey},
    number = {9},
    volume = {53},
    doi = {10.1109/TMAG.2017.2718485},
    issn = {00189464},
    arxivId = {arXiv:1701.03838v1}
}

\end{document}